\documentclass[runningheads]{llncs}
\usepackage[T1]{fontenc}
\usepackage{graphicx,verbatim}
\usepackage{graphicx}
\usepackage{amsfonts,amssymb}
\usepackage{caption}
\usepackage{color}
\usepackage{epstopdf}
\usepackage{tabularx,threeparttable}
\usepackage{array}
\usepackage[hidelinks]{hyperref}
\usepackage{graphicx}
\usepackage{subfigure}
\usepackage{amssymb, amsmath, bm}
\usepackage{mathtools}
\usepackage{mathrsfs}
\usepackage{booktabs}
\usepackage{multirow}
\usepackage{cite}
\usepackage{url}
\usepackage{color}
\usepackage{dsfont}
\usepackage{enumitem}
\usepackage[table,xcdraw]{xcolor}
\usepackage{hyperref}
\hypersetup{
    colorlinks = true,    
    citecolor = blue      
}

\usepackage{marvosym}

\begin{document}
\title{ADAgent: LLM Agent for Alzheimer's Disease Analysis with Collaborative Coordinator}
\titlerunning{ADAgent}
%
\authorrunning{W. Hou et al.}


\author{Wenlong Hou\textsuperscript{1}, Guangqian Yang\textsuperscript{1}, Ye Du\textsuperscript{1}, Yeung Lau\textsuperscript{1}, Lihao Liu\textsuperscript{2}, \\Junjun He\textsuperscript{3}, Ling Long\textsuperscript{4}, Shujun Wang\textsuperscript{1, 5(\Letter)}}  
\institute{\textsuperscript{1} Department of Biomedical Engineering, The Hong Kong Polytechnic University, Hong Kong SAR, China \\
\textsuperscript{2} Amazon, USA \\
\textsuperscript{3} Shanghai Artificial Intelligence Laboratory, China \\
\textsuperscript{4} Department of Geriatrics, Third Affiliated Hospital of Sun Yat-sen University, China \\
\textsuperscript{5} Research Institute for Smart Ageing, The Hong Kong Polytechnic University, Hong Kong SAR, China \\
\email{shu-jun.wang@polyu.edu.hk}}

\maketitle              

\begin{abstract}
Alzheimer’s disease (AD) is a progressive and irreversible neurodegenerative disease. Early and precise diagnosis of AD is crucial for timely intervention and treatment planning to alleviate the progressive neurodegeneration. 
However, most existing methods rely on single-modality data, which contrasts with the multifaceted approach used by medical experts. While some deep learning approaches process multi-modal data, they are limited to specific tasks with a small set of input modalities and cannot handle arbitrary combinations. This highlights the need for a system that can address diverse AD-related tasks, process multi-modal or missing input, and integrate multiple advanced methods for improved performance.
In this paper, we propose ADAgent, the first specialized AI agent for AD analysis, built on a large language model (LLM) to address user queries and support decision-making. ADAgent integrates a reasoning engine, specialized medical tools, and a collaborative outcome coordinator to facilitate multi-modal diagnosis and prognosis tasks in AD. Extensive experiments demonstrate that ADAgent outperforms SOTA methods, achieving significant improvements in accuracy, including a 2.7\% increase in multi-modal diagnosis, a 0.7\% improvement in multi-modal prognosis, and enhancements in MRI and PET diagnosis tasks.


\end{abstract}

\section{Introduction}
Alzheimer's disease (AD) is a neurodegenerative disorder characterized by the progressive decline in cognitive function, memory, and behavior \cite{AD_lancet,joe2019cognitive}. It affects millions of people worldwide, leading to a significant burden on patients, families, and healthcare systems. The impact of AD is not only detrimental to the patient’s quality of life but also has profound social and economic consequences \cite{better2023alzheimer,world2021global}. Early diagnosis plays a crucial role in slowing down the disease's progression and providing patients and families with opportunities for better care planning, treatment, and support \cite{early_diagnosis}. Traditional methods for diagnosing AD include cognitive tests, clinical assessments, and neuroimaging techniques such as magnetic resonance imaging (MRI) and positron emission tomography (PET) scans \cite{dubois2021clinical}. In the early stage of AD, subtle changes in the brain and mild cognitive decline would occur, which are also common in the normal ageing cases, leading difficulty to directly judge from any traditional methods mentioned above. In addition, the heterogeneity of data from different methods and the subjectivity in doctors' assessments increase the difficulty of extracting multi-modal information to identify early-stage AD patients.

With the recent advancements in deep learning (DL), new DL algorithms have emerged, offering more objective and efficient ways to automatically detect early AD instead of subjective assessments based on humans. Most methods~\cite{dl_AD1,dl_AD2,dl_AD3,dl_AD5,dl_AD101,dl_AD102,dl_AD6,dl_AD7,dl_AD8} in the AD domain rely on data from a single modality, which is contrary to the multifaceted approach that medical experts use to detect the disease. Despite several DL methods attempting to process multi-modal data~\cite{dl_AD4,dl_AD9,dl_AD10}, they remain limited in only handling a specific range of tasks with a small set of input modalities, and cannot tackle multiple tasks with arbitrary input modalities. In addition, even if multiple existing methods are designed for the same task, they cannot be collaboratively integrated to fully utilize their advantages for performance improvement. This highlights the need for a system with these three properties: (1) can address a wide range of AD-related tasks, (2) can process multi-modal input or missing modality input, and (3) can achieve objective assessments by collaborative coordinator of existing multiple methods. 

The emergence of AI Agents driven by LLMs has fundamentally changed how we approach autonomous reasoning, planning, and tool-using~\cite{medraxmedicalreasoningagent}. This paradigm shift has enabled AI Agents to surpass traditional task-specific models by dynamically adapting to diverse applications without additional training. There are several medical agents for medical reasoning on chest and abdomen X-rays~\cite{medraxmedicalreasoningagent, mmedagent}, showing notable performance on a wide range of tasks. Motivated by these, we aim to develop a solution via building an LLM-driven Agent integrated with multiple tools.


In this paper, we propose the Alzheimer's Disease Analysis Agent (ADAgent), the first specialized AI agent for AD analysis. Specifically, it is a large language model (LLM)-based system that could solve user (\textit{i.e. }doctors) queries (\textit{e.g.}, determining the disease stage) by incorporating reasoning, planning, and tool-using capabilities to facilitate task execution and support decision-making. ADAgent consists of three primary components: (1) a reasoning engine driven by the LLM, which serves as a tool action planner; (2) a set of specialized medical tools, each targeting specific tasks in AD detection; and (3) a collaborative outcome coordinator, which leverages the LLM’s reasoning capabilities to make informed decisions based on aggregated results. ADAgent is integrated with four tools capable of processing multi-modal input data for both the diagnosis and prognosis of AD. ADAgent can be extended to more tasks by accessing more tools/methods seamlessly without extra training. To access the performance of the proposed ADAgent, we extensively compared the result of ADAgent against multiple baseline methods on multi-modal tasks and missing modality diagnosis tasks. The results show that with the help of the collaborative outcome coordinator of ADAgent, the accuracy performance improves 2.7\%, 0.7\%, 0.7\%, and 4.4\% on multi-modal diagnosis task, multi-modal prognosis task, MRI diagnosis task, and PET diagnosis task, respectively.
    
    

\begin{figure}[t]
\includegraphics[width=\textwidth]{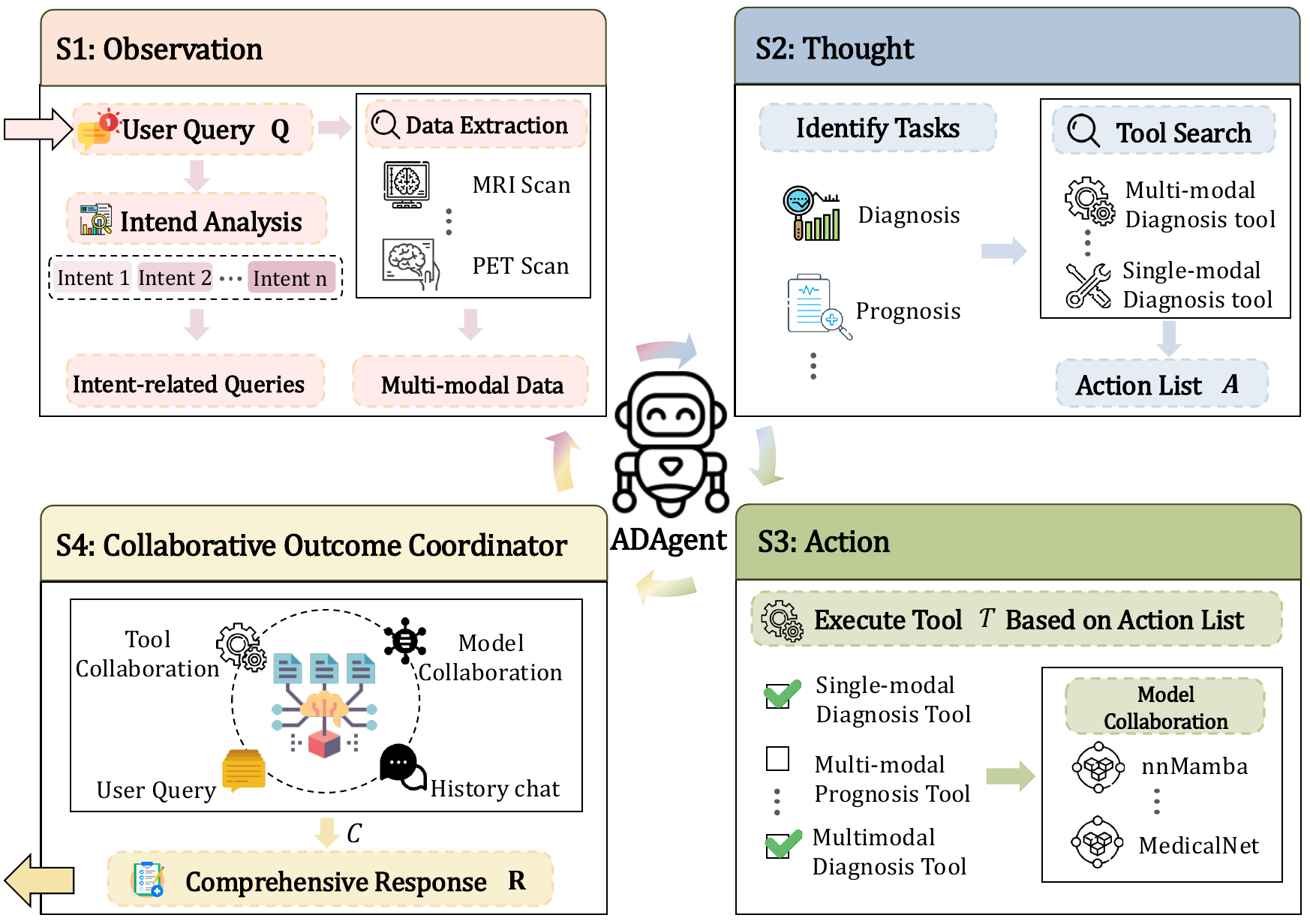}
\caption{Overall workflow of ADAgent. Given a use query, the framework performs reasoning in four stages: (i) observation, (ii) thought, (iii) action, and (iv) collaborative outcome coordinator.} \label{fig1}
\end{figure}

\section{ADAgent}

In this section, we first present the overall workflow of ADAgent in Section \ref{sec2.1}, followed by a detail description of all four modules in Section \ref{sec2.2}. The extensibility analysis and the interaction example of ADAgent are described in Section \ref{sec2.3} and \ref{sec2.4}, respectively.

\subsection{Overall Workflow of ADAgent}
\label{sec2.1}
Our developed ADAgent processes multi-modal use queries, including MRI scan and/or PET scan, along with text-based intents from users, to perform advanced analysis and generate diagnostic and prognostic results.
ADAgent employs a large language model (LLM), denoted by \( \mathcal{L} \), as the core to drive a reasoning and acting loop, which decomposes complex medical queries into sequential analytical steps. The loop is formalized as \( \mathcal{R} \) for reasoning and \( \mathcal{A} \) for acting, inspired by MedRAX approach \cite{medraxmedicalreasoningagent}, which is a medical reasoning agent for chest X-ray. To address a user query \( \mathbf{Q} \), ADAgent follows a four-step process, as shown in Fig.~\ref{fig1}: (1) \textbf{Observation} – The system observes the current state \( s \in \mathcal{S} \) and interprets the user’s query \( \mathbf{Q} \), where \( \mathcal{S} \) represents the state space. (2) \textbf{Thought} – The system determines the necessary actions \( \mathbf{A} \) to proceed, where \( \mathbf{A} \) is the action set.
(3) \textbf{Action} – The system executes relevant tools \( \mathcal{T}_i \) to gather outcomes \( \mathbf{O}_i \) from each tool, where \( \mathcal{T}_i \in \mathcal{T} \) are the tools in the tool set \( \mathcal{T} \).
(4) \textbf{Collaborative Outcome Coordinator} – The system synthesizes the findings from the previous steps to generate a comprehensive response \( \mathbf{R} \) for the user.

\subsection{\textbf{Observation}, \textbf{Thought}, \textbf{Action}, and \textbf{Collaborative Outcome Coordinator}}
\label{sec2.2}

\textbf{Observation Module -} The Observation module divides the user query into more refined sub-queries based on the user's intentions and then gathers the necessary text and image for the next Thought module.

\noindent \textbf{Thought Module with Tool Integration -}
ADAgent integrates a diverse set of tools \( \mathcal{T} = \{ \mathcal{T}_1, \mathcal{T}_2, \dots, \mathcal{T}_n \} \) designed to address various AD-related analysis tasks. For each tool \(\mathcal{T}_i\), we attempt to find multiple public models, denoted by \( \mathcal{M}_i = \{ \mathcal{M}_{i1}, \mathcal{M}_{i2}, \dots, \mathcal{M}_{ik} \} \), for collaboration to enhance task execution. ADAgent primarily focuses on two core tasks: diagnosis and prognosis.

\textbf{(i) Diagnosis task} aims to assess the current stage of AD based on user input, such as MRI or PET scan data. Formally, the diagnosis task can be expressed as
\begin{equation} 
    \hat{y}_{\text{diag}} = \mathcal{D}(\mathcal{X}, \mathcal{M}),
\end{equation}
where \( \hat{y}_{\text{diag}} \) represents the predicted stage of AD, \( \mathcal{X} = \{ \mathbf{X}_1, \mathbf{X}_2 \} \) denotes the multi-modal inputs (e.g., MRI and PET scans), and \( \mathcal{M} \) is the set of models used for diagnosis (e.g., CMViM, MCAD, etc.).

\textbf{(ii) Prognosis task} focuses on predicting the future progression of Alzheimer’s disease. The prognosis task can be modeled as
\begin{equation}
    \hat{y}_{\text{prog}} = \mathcal{P}(\mathcal{X}, \mathcal{M}),
\end{equation}
where \( \hat{y}_{\text{prog}} \) is the predicted progression, and \( \mathcal{X} \) and \( \mathcal{M} \) are the same as in the diagnosis task, but with models specialized for prognosis.

\textbf{(iii) Missing modality scenario} is common in clinical practice, where certain modalities may be unavailable or incomplete. In such cases, ADAgent integrates single-modal tools. If only one modality, say MRI \( \mathbf{X}_{\text{MRI}} \), is available, the system uses the MRI Diagnosis Tool \( \mathcal{T}_{\text{MRI}} \), and the prediction is made as
\begin{equation}
    \hat{y}_{\text{diag}} = \mathcal{T}_{\text{MRI}}(\mathbf{X}_{\text{MRI}}).
\end{equation}
Similarly, if only the PET modality is available, the PET Diagnosis Tool \( \mathcal{T}_{\text{PET}} \) is used to process \( \mathbf{X}_{\text{PET}} \) and make predictions.

Although ADAgent currently integrates existing tools for AD diagnosis and prognosis, its architecture is highly extensible to integrate new tools. Integrating a new tool requires only a class definition that specifies the tool’s input/output formats without any extra training, which are mentioned in detail in Section \ref{sec2.3}.

\noindent \textbf{Action Module -} The Action module follows the planning and executes the task by utilizing relevant tools to get necessary outcomes for further analysis.

\noindent \textbf{Collaborative Outcome Coordinator Module -}
The Collaborative Outcome Coordinator is responsible for synthesizing the response from the various specialized tools in ADAgent. Each tool \( \mathcal{T}_i \) is designed to handle a distinct AD-related analysis task, and the tool collaborates with multiple models to ensure accurate and efficient task execution. The collaborative outcome coordinator leverages the reasoning capabilities of the LLM \( \mathcal{L} \) to make informed decisions based on the aggregated results from the different tools. 
Mathematically, this process can be expressed as
\begin{equation}
    \mathbf{R} = \mathcal{C} \left( \mathcal{O}_1(\mathcal{M}_{1}), \mathcal{O}_2(\mathcal{M}_{2}), \dots, \mathcal{O}_n(\mathcal{M}_{n}) \right),
\end{equation}
where \( \mathcal{C} \) represents the coordinator function that combines the results from selected tools, and \(\mathcal{O}_i(\mathcal{M}_{i})\) denotes the output of tool \(\mathcal{T}_i\) based on \(\mathcal{M}_{i}\).
\begin{figure}[t]
\includegraphics[width=\textwidth]{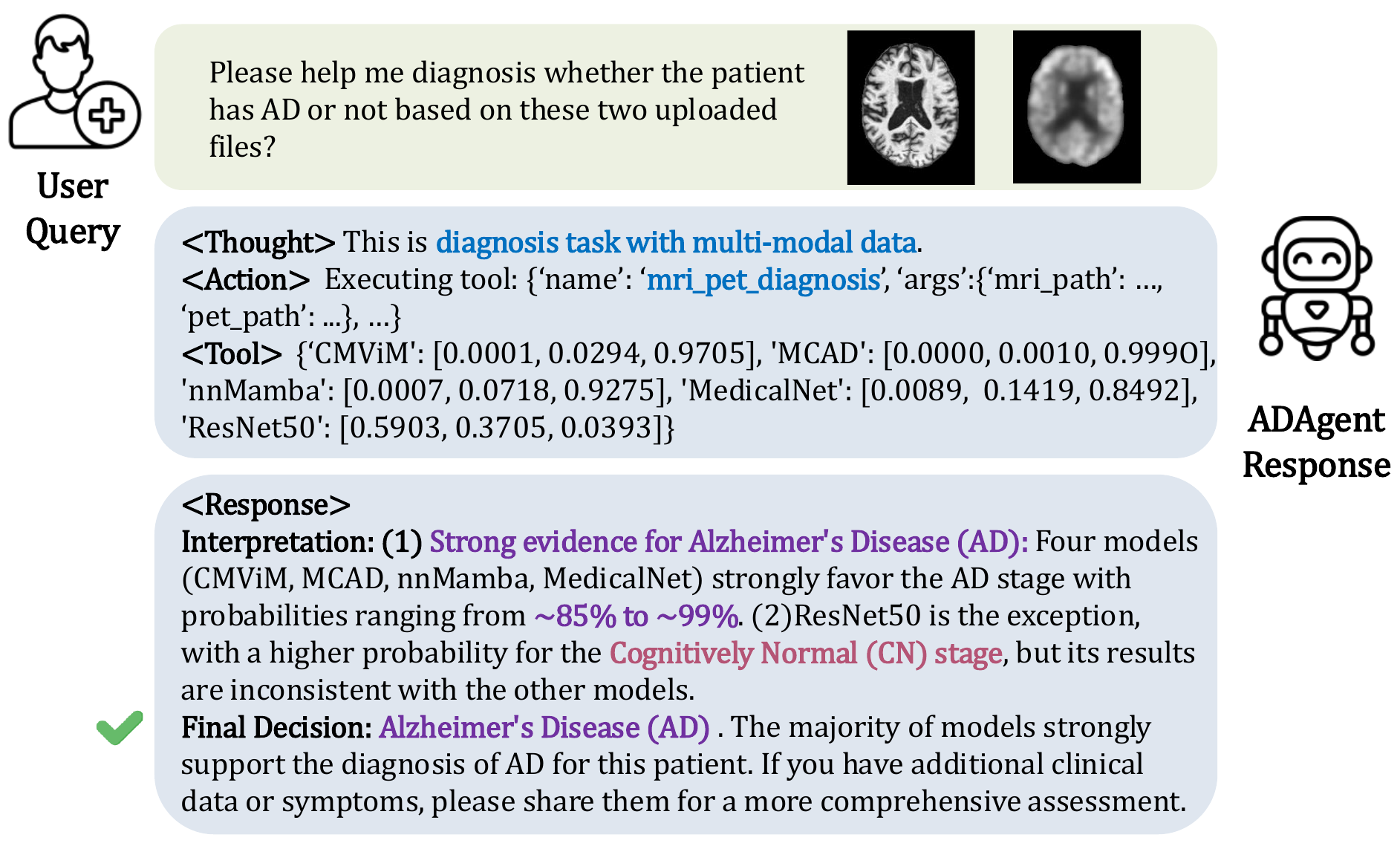}
\caption{ADAgent Interaction Example. An example of how
ADAgent handles a conversation through its workflow (<thought>, <action>, <Tool>) along with tool outputs and final response.} \label{fig2}
\end{figure}

\subsection{Extensibility}
\label{sec2.3}
ADAgent is built using the LangChain and LangGraph frameworks. Its reasoning engine can be powered by any LLM, supporting both text-only and multi-modal models, whether open-source or proprietary. This flexibility allows ADAgent to be deployed in various environments, from local installations to cloud-based solutions, while accommodating diverse healthcare privacy requirements~\cite{medraxmedicalreasoningagent}. Each tool in ADAgent functions as an independent module, with clearly defined input, output, and specific capability. The modular design allows tools to be modified, replaced, or repurposed without affecting the overall system. Integrating a new tool requires only a class definition that specifies the tool's input/output formats and capabilities, enabling the LLM to learn its usage without extra training.
\subsection{Interface and Interaction Example}
\label{sec2.4}
ADAgent includes a production-ready interface built with Gradio, facilitating seamless deployment in clinical settings. The interface supports the uploading of MRI and PET scans in NIFTI format and maintains an interactive chat session for natural interactions. The interface also provides transparency by tracking and displaying intermediate outputs during tool execution. An example interaction is shown in Fig.~\ref{fig2}.

\begin{table}[htbp]
	\centering
	\caption{Quantitative comparison on multi-modal tasks}
	\begin{tabular}{lcccc}
		\toprule  
		Method&ACC&SPE&SEN&F1\\ 
		\cmidrule(r){1-5}
            \rowcolor{gray!20}
            \multicolumn{5}{c}{\textit{Multi-modal diagnosis task}}\\
		MedicalNet~\cite{MedicalNet}&0.571±0.009&0.768±0.009&0.591±0.030&0.586±0.015\\
            nnMamba~\cite{nnmamba}&0.527±0.016&0.753±0.008&0.572±0.021&0.541±0.028\\
            ResNet50~\cite{ResNet}&0.567±0.039&0.771±0.018&0.599±0.036&0.581±0.043\\
            MCAD~\cite{MCAD}&0.548±0.029&0.766±0.014&0.591±0.027&0.556±0.026\\
            CMViM~\cite{dl_AD8}&0.617±0.014&0.659±0.003&0.630±0.013&0.633±0.010\\
            \cmidrule(r){1-5}
            \textbf{ADAgent (Ours)}&\textbf{0.644±0.014}&\textbf{0.794±0.010}&\textbf{0.644±0.021}&\textbf{0.647±0.018}\\
            \toprule
            \rowcolor{gray!20}
            \multicolumn{5}{c}{\textit{Multi-modal prognosis task}}\\
            MedicalNet~\cite{MedicalNet}&0.763±0.026&0.902±0.017&0.333±0.052&0.407±0.064\\
            nnMamba~\cite{nnmamba}&0.733±0.038&0.843±0.068&0.394±0.189&0.405±0.147\\
            ResNet50~\cite{ResNet}&0.726±0.034&0.853±0.029&0.333±0.052&0.374±0.065\\
            MCAD~\cite{MCAD}&0.719±0.026&0.892±0.061&0.182±0.091&0.231±0.080\\
            CMViM~\cite{dl_AD8}&0.815±0.046&0.912±0.029&0.515±0.105&\textbf{0.575±0.109}\\
            \cmidrule(r){1-5}
            \textbf{ADAgent (Ours)}&\textbf{0.822±0.010}&\textbf{0.941±0.019}&\textbf{0.545±0.043}&0.561±0.030\\
		\bottomrule  
        \label{multi-modal task}
	\end{tabular}
\end{table}
\section{Experiments}
\subsection{Dataset}
This study was conducted on the Alzheimer’s Disease Neuroimaging Initiative (ADNI) dataset\cite{ADNI}, and focused on T1-MRI and PET-FDG scans. As for the diagnosis task, we aim to determine the current stage of AD process based on use input from three stages: Cognitively normal (CN), Mild Cognitive Impairment (MCI), and AD. The prognosis task is to predict whether an MCI patient will convert to AD within 36 months or remain stable. ADAgent can directly integrate existing models without extra training. However, since there are few open-source models, we trained some models for integration in this work. 

\begin{table}[htbp]
	\centering
	\caption{Quantitative comparison on missing modality diagnosis tasks}
	\begin{tabular}{lcccc}
		\toprule  
		Method&ACC&SPE&SEN&F1\\ 
		\cmidrule(r){1-5}
            \rowcolor{gray!20}
            \multicolumn{5}{c}{\textit{Diagnosis with MRI}}\\
		MedicalNet~\cite{MedicalNet}&0.521±0.022&0.760±0.004& 0.569±0.007& 0.553±0.024\\
            ResNet50~\cite{ResNet}&0.536±0.012&\textbf{0.762±0.002}&\textbf{0.571±0.004}&\textbf{0.565±0.014}\\
            ResNet34~\cite{ResNet}&0.487±0.006&0.726±0.008&0.504±0.026&0.479±0.027\\
            ResNet18~\cite{ResNet}&0.515±0.016&0.739±0.010&0.533±0.024&0.522±0.016\\      
            \cmidrule(r){1-5}
            \textbf{ADAgent (Ours)}&\textbf{0.543±0.007}&0.754±0.003&0.562±0.013&0.553±0.011\\
            \toprule
            \rowcolor{gray!20}
            \multicolumn{5}{c}{\textit{Diagnosis with PET}}\\
            MedicalNet~\cite{MedicalNet}&0.550±0.022&0.760±0.004&0.569±0.007&0.553±0.024\\
            ResNet50~\cite{ResNet}&0.550±0.035&0.755±0.014&0.554±0.030&0.554±0.040\\
            ResNet34~\cite{ResNet}&0.524±0.028&0.751±0.008&0.555±0.014&0.523±0.028\\
            ResNet18~\cite{ResNet}&0.520±0.015&0.747±0.010&0.552±0.031&0.529±0.022\\        
            \cmidrule(r){1-5}
            \textbf{ADAgent (Ours)}&\textbf{0.594±0.034}&\textbf{0.785±0.014}&\textbf{0.631±0.027}&\textbf{0.611±0.034}\\
		\bottomrule  
        \label{missing modality task}
	\end{tabular}
\end{table}

\subsection{Experimental Settings}
ADAgent uses GPT-4o as its backbone LLM and is deployed on a single NVIDIA RTX 4090 GPU. ADAgent integrates four tools: (1) multi-modal diagnosis tool, and (2) multi-modal prognosis tool, both integrating five models: MedicalNet, nnMamba, ResNet50, MCAD, CMViM; (3) MRI diagnosis tool, and (4) PET diagnosis tool, both embedding MedicalNet, ResNet50, ResNet34 and ResNet18 for missing modality tasks. All experiments are conducted three times, and the average and standard deviation are reported for analysis.
\subsection{Experimental Results} 
To access the performance of the proposed ADAgent, we extensively compared the result of ADAgent against multiple baseline methods on multi-modal tasks and missing modality diagnosis tasks. To ascertain whether the performance improvement is attributable to the LLM's capability to aggregate the results of multiple models, we conducted the ablation study on multi-modal tasks and missing modality diagnosis tasks against two baseline methods.

\begin{table}[htbp]
	\centering
	\caption{Ablation study on ADAgent}
	\begin{tabular}{lcccc|cccc}
		\toprule  
		Method&ACC&SPE&SEN&F1&ACC&SPE&SEN&F1\\ 
            \cmidrule(r){1-9}
            \rowcolor{gray!20}
            \textit{Multi-modality}&\multicolumn{4}{c}{\textit{Diagnosis task}}&\multicolumn{4}{c}{\textit{Prognosis task}}\\
            Average&0.593&0.786&0.631&0.613&0.770&0.931&0.273&0.355\\
            Vote&0.575&0.778&0.615&0.589&0.763&0.941&0.212&0.301\\
            \cmidrule(r){1-9}
            \textbf{ADAgent (Ours)}&\textbf{0.644}&\textbf{0.794}&\textbf{0.644}&\textbf{0.647}&\textbf{0.822}&\textbf{0.941}&\textbf{0.545}&\textbf{0.561}\\
            \toprule
            \rowcolor{gray!20}
            \textit{Missing Modality}&\multicolumn{4}{c}{\textit{MRI diagnosis task}}&\multicolumn{4}{c}{\textit{PET diagnosis task}}\\		Average&0.519&0.743&0.546&0.530&0.582&0.782&0.623&0.595\\            Vote&0.536&\textbf{0.757}&\textbf{0.570}&0.546&0.582&0.782&0.619&0.595\\    
            \cmidrule(r){1-9}
            \textbf{ADAgent (Ours)}&\textbf{0.543}&0.754&0.561&\textbf{0.553}&\textbf{0.644}&\textbf{0.794}&\textbf{0.644}&\textbf{0.647}\\            
		\bottomrule  
        \label{abl_study}
	\end{tabular}
\end{table}

\textbf{Results on Multi-modal Tasks.}
Table \ref{multi-modal task} presents the result comparisons of ADAgent against five baselines: MedicalNet, nnMamba, ResNet50, MCAD, and CMViM, in terms of Accuracy (ACC), Specificity (SPE), Sensitivity (SEN), and F1-score (F1) on multi-modal tasks. ADAgent outperforms baseline methods on accuracy by 7.3\%, 11.7\%, 7.7\%, 9.6\%, and 2.7\% for the diagnosis task, and by 5.9\%, 8.9\%, 9.6\%, 10.3\%, and 0.7\% for the prognosis task. Especially, ADAgent achieves the best results in almost all metrics for these two tasks. These results validate ADAgent's reasoning ability in analyzing different models' results and can indeed improve performance.

\textbf{Results on Missing Modality Diagnosis Tasks.}
Similarly, Table \ref{missing modality task} compares the ADAgent against four baseline approaches, MedicalNet, ResNet50, ResNet34, and ResNet18 on missing modality diagnosis tasks. As for the accuracy metric, ADAgent outperforms baseline methods by 2.2\%, 0.7\%, 5.6\%, and 2.8\% for MRI diagnosis tools and by 4.4\%, 4.4\%, 7.0\%, and 7.4\% for the PET diagnosis tool. It is noted that although ResNet34 demonstrates poor performance in MRI diagnosis tool, ADAgent still has accuracy improvement and retains relatively good performance in other metrics. By contrast, ADAgent achieves the best results in all metrics on the PET diagnosis task. These results show that ADAgent has reasoning ability when integrating the results of different models, bringing performance improvement and robustness.

\textbf{Ablation Study.}
 We conducted the ablation study on all tasks against two baseline methods, and the detailed results are shown in Table \ref{abl_study}. The Average method is to average the prediction probabilities for all classes and select the class with the largest probability as the final prediction. The Vote method involves each model providing a decision, and the result that occurs most frequently is taken as the final outcome. ADAgent yields the best overall performance on all tasks. This suggests that ADAgent can fully leverage the capability of LLM to analyze the results of multiple models.

\section{Conclusion} 
In conclusion, ADAgent introduces the first AI agent framework integrated with multiple tools for AD analysis, powered by an LLM as both a tool action planner and a collaborative outcomes coordinator. Experimental results demonstrate that ADAgent outperforms existing methods, showcasing its potential for advanced AD diagnosis and prognosis. However, the current implementation of ADAgent is limited by the number of integrated tools and input modalities. Future work will focus on expanding the tool set, incorporating additional modalities, and enhancing the model’s generalization to further improve its clinical applicability.

%
%
%

\bibliographystyle{splncs04}
\bibliography{ref}

\end{document}